\documentclass[12pt,tcilatex]{article}

\usepackage{amsmath}
\input{tcilatex}

\date{}
\begin{document}

\title{\large Adiabatic waves along interfacial layers  near the critical
point\\\vskip 0.25 cm {\large {\it Ondes adiabatiques le long des
interfaces pr\`es du point critique}}}
\author{ Henri Gouin \\
{\small {Laboratoire de Mod{\'e}lisation en  M{\'e}canique et
Thermodynamique,
EA 2596} }\\
{\small {Universit{\'e} d'Aix-Marseille, 13397 Marseille Cedex 20,
France}}} \maketitle

\begin{abstract}
Near the critical point, isothermal interfacial zones are
investigated starting from a non-local density of energy.   From
the equations of motion of {\it thermocapillary fluids}
\cite{casal}, we point out a new kind of adiabatic waves
propagating along the interfacial layers. The waves are associated
with the  second derivatives of densities and propagate with a
celerity depending on the proximity of the critical point.\\

\centerline{\bf R\'esum\'e} \vskip 0.25 cm {\footnotesize  {\it
Pr\`es du point critique, les couches interfaciales sont
mod\'elis\'ees \`a l'aide d'une densit\'e d'\'energie non locale.
A partir des \'equations du mouvement des} fluides
thermocapillaires \cite{casal}, {\it nous mettons en \'evidence
des ondes adiabatiques   se propageant le long des couches
interfaciales. Ces ondes associ\'ees aux d\'eriv\'ees secondes des
densit\'es se meuvent avec une c\'el\'erit\'e d\'ependant de la
proximit\'e du point critique.}}

\end{abstract}
\vskip 0.7cm

{\noindent {\bf \large Version fran\c caise abr\'eg\'ee}}

\vskip 0.7cm
 Le
mod\`ele le plus simple permettant de consid\'erer les couches
capillaires et les phases comme un unique milieu continu, consiste
\`a  prendre en compte une \'energie  comme  la somme de deux
termes : le premier correspond \`a l'\'energie du milieu suppos\'e
uniforme et de composition \'egale \`a la composition locale, le
second est associ\'e \`a la non uniformit\'e du fluide et
exprim\'e par un d\'eveloppement en gradient
 de la masse volumique qui est limit\'e au second
ordre \cite{vdW,cahn}. Ce mod\`ele permet  le prolongement au cas
dynamique des \'etudes effectu\'ees sur les syst\`emes en
\'equilibre. L'\'energie interne volumique du fluide est
maintenant propos\'ee sous la forme d'une densit\'e $\,\alpha\,$
d\'ependant non seulement de grad$\,\rho\,$ mais aussi de
grad$\,s\,$ o\`u $\rho$ et $s$ notent respectivement la masse
volumique et l'entropie sp\'ecifique. Les milieux associ\'es
appel\'es {\it fluides thermocapillaires} \cite{casal} ont une
densit\'e d'\'energie de la forme (\ref{A}). Pour un fluide
isotrope, l'\'energie prend la forme (\ref{B}) et l'\'equation du
mouvement s'\'ecrit sous les formes \'equivalentes (\ref{G}) ou
(\ref{GG}) dans lesquelles   une entropie $h$ et une temp\'erature
$\theta$ dites {\it thermocapillaires}\, explicit\'ees en
(\ref{E}) font intervenir deux nouveaux vecteurs
${\mbox{{\boldmath $\Phi$}}} $ and ${\mbox{{\boldmath $\Psi$}}} $
associ\'es \`a la non homog\'en\'eit\'e des zones interfaciales et
donn\'es par (\ref{D}). Une telle \'energie interne associe dans
les \'equations du mouvement et de l'\'energie un tenseur des
contraintes sph\'erique. Le cas des \'ecoulements isentropiques
associ\'es \`a l'\'equation (\ref{GH}) correspond \`a la
conservation de l'\'energie (\ref{GK}). Les mouvements isothermes
repr\'esent\'es par (\ref{H1}) et (\ref{H2}) ont comme cas
particulier les \'equilibres isothermes entre phases. \\
Il est alors
 possible d'\'etudier les profils des densit\'es dans
les \'equilibres unidimensionnels correspondant aux zones
interfaciales {\it planes}. Le choix d'une \'energie interne
thermocapillaire de la forme (\ref{I}) associ\'ee \`a une
\'energie interne $\alpha$ du milieu suppos\'e homog\`ene ram\`ene
l'\'etude des zones interfaciales \`a l'analyse d'un syst\`eme
dynamique (\ref{F1}), (\ref{F2}) qui donne deux \'equations
diff\'erentielles. L'analyse asymptotique de ce syst\`eme au
voisinage du point critique montre que la densit\'e de masse {\it
pilote} le syst\`eme en ce sens que l'on est ramen\'e \`a une
d\'ecomposition dynamique repr\'esent\'ee par le syst\`eme
(\ref{J4}), (\ref{KK}). L'int\'egration de ce syst\`eme nous
ram\`ene simplement au cas d'un mod\`ele ne faisant plus
intervenir que grad$\, \rho$. On retrouve imm\'ediatement dans
(\ref{K5}), les r\'esultats propos\'es par Rowlinson et Widom
\cite{rowlinson} pour le profil de la masse volumique,
l'\'epaisseur de la couche interfaciale et la valeur de la tension
superficielle des
interfaces fluides au voisinage du point critique.\\
Il est maintenant
 possible d'\'etudier, dans les interfaces
isothermes, les ondes d'acc\'el\'eration correspondant \`a des
discontinuit\'es faibles des mouvements isentropiques
c'est-\`a-dire celles  pour lequelles $\rho, s,$ grad$\,\rho$ et
grad$\,s$ sont continus \`a la travers\'ee des surfaces d'onde. Un
calcul {\it \`a la Hadamard} \cite{hadamard} est propos\'e. Il
prend en compte l'\'equation de conservation de la masse
(\ref{K}), l'\'equation du mouvement sous la forme (\ref{G}) et la
condition de Rankine-Hugoniot associ\'ee \`a l'\'equation du
mouvement \'ecrite sous la forme \'equivalente (\ref{GG}). Il
permet alors d'obtenir un syst\`eme (\ref{Q}), (\ref{R2}),
(\ref{R3}) de trois \'equations lin\'eaires et homog\`enes relatif
aux discontinuit\'es des d\'eriv\'ees secondes \`a travers les
surfaces d'onde (paragraphe 4.1). La condition de compatibilit\'e
de ce syst\`eme s'exprime par l'\'equation
(\ref{T}) qui donne ainsi la c\'el\'erit\'e de l'onde.\\
La relation (\ref{KK}) issue du syst\`eme dynamique  (\ref{KY}),
(\ref{KZ}) obtenu par l'ana-lyse asymptotique du paragraphe 3,
permet d'expliciter la relation liant les gradients d'entropie et
de masse volumique. Nous la calculons  pour la valeur de la masse
volumique critique. Nous en d\'eduisons la valeur explicite de la
c\'el\'erit\'e des ondes d'acc\'el\'eration cisaill\'ees dans les
interfaces fluides. Cette c\'el\'erit\'e exprim\'ee par la
relation (\ref{U}) d\'epend des conditions thermodyna-miques du
point critique et de l'importance des coefficients associ\'es aux
termes repr\'esentant l'inhomog\'en\'eit\'e du fluide dans la zone
interfaciale. Elle est proportionnelle \`a $(T_c-T_0)^2$ o\`u
$T_0$ repr\'esente la valeur de la temp\'erature dans les phases
liquide et vapeur et $T_c$ note la temp\'erature
critique.\\
 Les ondes solitaires dans la direction
normale aux zones interfaciales ne d\'ependent pas du gradient
d'entropie. Il n'en est pas de m\^eme des ondes isentropiques
d'acc\'el\'eration qui se meuvent le long des interfaces : le fait
que l'\'energie interne d\'epend non seulement du gradient de
masse volumique mais aussi du gradient d'entropie fait
appara\^{i}tre un nouveau type d'onde qu'il est impossible de
mettre en \'evidence dans les mod\`eles plus simples ne faisant
intervenir que le gradient de masse volumique. Ces ondes
exceptionnelles au sens de Lax \cite{Ruggeri} n\'ecessitent le
cadre d'un mod\`ele physique \`a au moins deux  dimensions
d'espace. Des exp\'eriences effectu\'ees r\'ecemment dans des
laboratoires embarqu\'es dans des engins spatiaux et utilisant
comme fluide test le gaz carbonique dans des conditions qui le
placent au voisinage de son point critique semblent montrer
l'existence de telles ondes \cite{Garrabos} et justifieraient
alors le bien fond\'e du mod\`ele de fluides thermocapillaires.

\section{Introduction}
To study capillary layers and bulk phases, the simplest model
considers an energy as the sum of two terms:  a first one
corresponding to a medium with a uniform composition equal to the
local one and a second one associated with the non-uniformity of
the fluid \cite{vdW,cahn}. The second term is approximated by a
gradient expansion, typically truncated to the second order. A
representation of the energy near the critical point therefore
allows the study of interfaces of non-molecular size. Obviously,
the model is simpler than models associated with the
renormalization-group theory \cite{domb}. Nevertheless, it has the
advantage of extending easily well-known results for equilibrium
cases to the dynamics of interfaces \cite{slemrod,trusk}. For
equilibrium, Rowlinson and Widom \cite{rowlinson} pointed out that
the model can be extended by taking into account not only the
strong variations of matter density through the interfacial layer
but also the strong variations of entropy. In dynamics, for an
{\it extended Cahn and Hilliard fluid}, the volumic internal
energy $\varepsilon$ is proposed with a gradient expansion
depending not only on $ {\rm grad}\ \rho\, $ but also on ${ \rm
grad}\ s $ ($\rho$ is the matter density and $s$ the specific
entropy) :
\begin{equation}
\varepsilon = f(\rho, s, {\rm grad}\ \rho,   {\rm grad}\ s ).
\label{A}
\end{equation}
The medium is then called a {\it thermocapillary fluid}
\cite{casal}. Using  an energy in the   form (\ref{A}), we have
 obtained the
equations  of conservative motions for nonhomogeneous fluid near
its critical point \cite{casal,gouin0,casal1}. \\ The idea of
studying interface motions as localized travelling waves in a
multi-gradient theory is not new and can be traced throughout many
problems of condensed matter and phase-transition physics
\cite{gouin2}. In Cahn and Hilliard's model \cite{cahn}, the
direction of solitary waves was along the gradient of density
\cite{slemrod,gouin2}. Here, adiabatic waves are considered and a
new kind of waves appears. The waves are associated with the
spatial second derivatives of entropy and matter density.  For
this new kind of adiabatic waves, the direction of propagation is
normal to the gradient of densities. In the case of a thick
interface, the waves are tangential to the interface and the wave
celerity is expressed
depending on thermodynamic conditions at the critical point.\\

\section{Equations of thermocapillary fluid motions}
The equations of motion are proposed in
\cite{casal,gouin0,casal1}.  Due to the fact the fluid is
isotropic, ${\rm grad}\ \rho$ and ${\rm grad}\ s$ are taken into
account by their scalar products only. Let us denote
$$
\beta = ({\rm grad}\ \rho)^2, \ \chi = {\rm grad}\ \rho\,.\, {\rm
grad}\ s, \  \gamma = ({\rm grad}\ s)^2.
$$
In variables $\rho, s, \beta, \chi, \gamma,$
\begin{equation}
\varepsilon = g(\rho, s, \beta, \chi, \gamma).\label{B}
\end{equation}
The equations of thermocapillary fluids introduced two new vectors
${\mbox{{\boldmath $\Phi$}}} $ and ${\mbox{{\boldmath $\Psi$}}} $
such that :
 \begin{equation}
 {\mbox{{\boldmath
$\Phi$}}} = C\ {\rm grad}\ \rho + D\ {\rm grad}\ s, \hskip 0.5cm
 {\mbox{{\boldmath
$\Psi$}}} = D\ {\rm grad}\ \rho + E\ {\rm grad}\  s, \label{D}
  \end{equation}
  with
  $$ C = 2 \ \varepsilon_{,\beta},\ D = \varepsilon_{,\chi}, \ E =
  2\ \varepsilon_{,\gamma}.
  $$
  We denote
\begin{equation}
 h = \varepsilon_{,\rho}\ - {\rm div}\
{\mbox{{\boldmath $\Phi$}}}, \hskip 0.5cm  \theta = {1\over \rho}
\left( \varepsilon_{,s} - {\rm div}\ {\mbox{{\boldmath
$\Psi$}}}\right). \label{E}
\end{equation}
  In the case of compressible fluids, scalars $\varepsilon_{,\rho}$
and $(1/\rho)\, \varepsilon_{,s}$ are the specific enthalpy and
the Kelvin temperature. Look at two particular cases :
\subsection{ Conservative  motions }
 \noindent   We obtained \cite{casal} the equation of motion in the form :
\begin{equation}
{\mbox{{\boldmath $\Gamma$}}} = \theta \ {\rm grad}\, s - {\rm
grad}(h+\Omega), \label{G}
\end{equation}
where ${\mbox{{\boldmath $\Gamma$}}}$ is the acceleration vector,
$\Omega$ is the extraneous force potential.  This equation  is
equivalent to the balance of momentum:
\begin{equation}
{\partial \over\partial t}\,(\rho {\mathbf u}) + {\rm div} (\rho\,
{\mathbf u} \otimes {\mathbf u}-\,\sigma) + \rho\, {\rm grad}\,
\Omega\ =\, 0 \label{GG}
\end{equation}
with $ \displaystyle \sigma_i^j = - ( P - \rho\, {\rm div}\,
{\mbox{{\boldmath $\Phi$}}})\, \delta_i^j - \Phi^j\,\rho_{,i}-
\Psi^j\,s_{,i} \, $  where  $\, P = \rho\, \varepsilon_{,\rho}
-\varepsilon$ (in the case of classical  compressible fluids, $P$
denotes the pressure) and ${\mathbf{u}}$ is the fluid velocity.
For conservative motions,
\begin{equation}
  {ds\over dt} = 0, \label{GH}
\end{equation}
which is equivalent to
 the balance of energy
\begin{equation}
{\partial e\over\partial t}\,  + {\rm div} \big((e - \sigma)
{\mathbf u} \big) -{\rm div}\, {\mathbf W} - \rho\, {\partial
\Omega\over\partial t} =\, 0, \label{GK}
\end{equation}
with $ \displaystyle {\mathbf W} = { {d\rho\over dt}}\,
{\mbox{{\boldmath $\Phi$}}} +  {ds\over dt}\, {\mbox{{\boldmath
$\Psi$}}} \, $ and $\, e =  {1\over 2}\ \rho {\mathbf u}^2
 + \varepsilon + \rho \Omega.$
\subsection{ Isothermal  motions }
\noindent We obtained \cite{casal} the equation of motion in the
form :
\begin{equation}
\theta = T_0,\label{H1}
\end{equation}
\begin{equation}
{\mbox{{\boldmath $\Gamma$}}} =  -\ {\rm grad}(\mu +\Omega),
\label{H2}
\end{equation}
where $T_0$ is constant and $ \mu = \varepsilon_{,\rho} - s\, T_0-
{\rm div}\ {\mbox{{\boldmath $\Phi$}}}$ is the {\it chemical
potential} of the thermocapillary fluid.

\section{Liquid-vapor interface near its critical point}
The critical point associated with the equilibrium of two bulks of
a fluid corresponds to   the limit of their coexistence. The
thickness of the interface increases as the critical point is
approached and it becomes infinite when the interface itself
disappears.   As its critical point is approached, the gradients
of densities are then smooth.  In the following, we consider the
 case when
\begin{equation}
\varepsilon = \rho\, \alpha(\rho,s) + {1\over 2}\Big ( C\ ({\rm
grad}\ \rho)^2 + 2 D\   {\rm grad}\ \rho\, . \, {\rm grad}\ s   +
E\ ({\rm grad}\ s)^2 \Big ) \label{I}
\end{equation}
where $\alpha$ denotes the specific internal energy of the fluid
in uniform composition (let $\alpha (\rho,s)$ be the analytic
$\alpha$ as it might be given by a mean-field theory), $C, D, E$
are constants and $C E - D^2
> 0$.  If $ D = 0$ and $ E = 0$, we are back to the Cahn and
Hilliard model of capillarity \cite{cahn}. If not, we deduce
$$
 h = h_0 - (C \ \Delta \, \rho + D\ \Delta \, s),
\hskip 0.5cm \theta =  T - {1\over \rho}\ ( D \, \Delta \ \rho +
E\ \Delta \, s),
$$
where  $h_0 \equiv \alpha + \rho\,\alpha^\prime_\rho  $ and
$T\equiv \alpha^\prime_s$  are respectively  the specific enthalpy
and
 the Kelvin temperature of the homogeneous fluid of matter density $\rho$
and specific entropy $s$; we denote by  $\mu_0 =
\varepsilon^\prime_\rho - s\, T $
its chemical potential.\\
\noindent At phase equilibrium, Eq. (\ref{H1}) is verified when
$T_0$ is the temperature in the bulks. If we neglect the body
forces, we obtain
\begin{subequations}
\begin{eqnarray}
D \ \Delta \, \rho + E\ \Delta \, s &=& \varepsilon^\prime_s\ -
\rho\, T_0\, ,\label{F2} \\
C \ \Delta \, \rho + D\ \Delta \, s &=&
\varepsilon^\prime_\rho - s\, T_0  - \mu_1\label{F1}
\end{eqnarray}
\end{subequations}
where $\mu_1 $ is  the value of $\mu_0$ in the liquid and the
vapor bulks.\\ To be in accordance with Rowlinson and Widom
(\cite{rowlinson} p. 253), we consider a representation of $\rho\,
\alpha(\rho,s)$ in the form
\begin{equation}
\rho\, \alpha(\rho,s)= {B\over 2 A^2} \left(\Big(A(\rho-\rho_c)^2
+ \rho s\Big)^2+ \Big(\rho s+ {A^2\over B} (T_c-T_0)\Big)^2
\right)+ \rho\mu_1 +\rho s T_0 \label{J3}
\end{equation}
where A and B are two positive constants associated with the
critical conditions  and $s$ which is defined except to any
additive constant is choosen such that $s_c = 0$. System
(12) yields\\

$ \displaystyle D\ \Delta\, \rho + E \ \Delta\, s = {B\over A}\,
\rho\, (\rho-\rho_c)^2 + 2\, {B\over A^2}\, \rho^2\,
s + \rho\,(T_c-T_0),$\\

$\displaystyle  C\ \Delta\, \rho + D\ \Delta\, s = 2\, B\,
(\rho-\rho_c)^3 +2\,{B\over A}\, \rho\, s\, (\rho-\rho_c) +
{B\over A} (\rho-\rho_c)^2\, s $

$\hskip 3 cm \displaystyle +\, 2\,{B\over A^2}\, \rho\, s^2 +
s\,(T_c-T_0)\,.\\ $

\noindent The equations of the coexistence curve at $T = T_0$ are
associated with $\Delta\, \rho =0 $ and $ \Delta\, s =0$. They are
as in \cite{rowlinson}, \, $\displaystyle A (\rho-\rho_c)^2 + \rho
s =0 $ and $ \displaystyle \rho s + {A^2\over B}(T_c-T_0)=0$.\\ In
the bulks, near the critical point,  the respective magnitudes of
$\rho-\rho_c$ and $ s $ with respect to $T_c-T_0$ are $\rho-\rho_c
\sim cte \,(T_c-T_0)^{1\over 2},\   s \sim cte \, (T_c-T_0).$
 In
one-dimensional problems, $\rho = \rho(y)$, $s = s(y)$. To
consider the respective order of magnitude of the densities and
the physical scales associated with the interfacial sizes, we look
at the change of variables
$$
T_c - T_0 = \epsilon \Upsilon,  \ y = \epsilon^n\, Y,\
\rho(y)-\rho_c = \epsilon^{1\over 2}\, R(Y) ,\ \ s(y) = \epsilon\,
S(Y),
$$
where $0 < \epsilon \ll 1$, and $\, n$ is constant. The main part
of system (12) leads to
 \begin{subequations}
\begin{eqnarray}
\epsilon^{-2n}\left( D\ \epsilon^{1\over 2}\, { d^2 R\over dY^2} +
E \ \epsilon  { d^2S\over dY^2}\right)&=& \epsilon \left( {B\over
A}\, \rho\, R^2 +  2\, {B\over A^2}\, \rho^2\, S + \Upsilon
\rho\right)\label{KZ}\\
 \epsilon^{-2n}\left( C\ \epsilon^{1\over
2}\, { d^2 R\over dY^2} +
 D \ \epsilon \,  { d^2S\over dY^2}\right) &=& \epsilon^{3\over 2}\left( 2\,
 B\, \,R^3 +  2\,{B\over A}\, \rho\,R\, S\right)\, \label{KY}
\end{eqnarray}
\end{subequations}
Then $n =-{1\over 2}$ and in this asymptotic analysis, if
$\,^{\prime\prime}$ denotes the second derivative with respect to
the space variable $y$, system (12) yields the approximation
\begin{subequations}
\begin{eqnarray}
2\ \rho s &=& {A^2\over B}\ (T_0 -T_c) - A\
(\rho-\rho_c)^2,\label{KK}\\
C\ \rho^{\prime \prime}  &=& 2\, B\, (\rho-\rho_c)^3 +2\,{B\over
A}\,
 \rho s\, (\rho-\rho_c)\label{J4}
\end{eqnarray}
\end{subequations}
and consequently by elimination of $s$ between (\ref{J4}) and
(\ref{KK}), we obtain
\begin{equation}
C\ \rho^{\prime \prime}  =  B\ (\rho-\rho_c)^3 - A\ (T_c
-T_0)(\rho-\rho_c),\label{K5}
\end{equation}
and for planar liquid-vapor interfaces,
\begin{equation}
{1\over 2}\ C \rho^{\prime 2}= \left( {{\sqrt B}\,\over
2}(\rho-\rho_c)^2 - { A\over 2 {\sqrt B}}\, (T_c -T_0)\right)^2.
\label{L}
\end{equation}
 We are back to the
classical van der Waals theory for mass density profile in the
interfacial layer (see \cite{rowlinson}, p.p. 250-251).

\section{Weak discontinuity in conservative motions}
\subsection{Conditions of a weak discontinuity}
In a fixed coordinate system,  to describe the fluid motion, we
refer to the coordinates $\mathbf{x} \equiv\ (x^1,x^2,x^3)$ as the
particle position (Eulerian variables). We denote by
 $ {\mathbf z} =$
 $ {\left(\begin{array}{c} t \\{\mathbf x} \end{array} \right)}$
the time-space variables;  ${\mathbf{V}} \equiv
\left(\begin{array}{c} 1 \\ {\mathbf u}
\end{array} \right)$
is the time-space  velocity. Let us consider a mobile surface
$\Sigma_t$ defined in the physical space occupied by the fluid, we
denote by $g$  the celerity of $\Sigma_t, \ {\mathbf n}$ its
normal vector,  $\mathbf{N} = \left(\begin{array}{c} -g \\
{\mathbf n}
\end{array} \right)$;
$v = {\mathbf{N}}^* {\mathbf{V}} \equiv
{\mathbf{n}}^*{\mathbf{u}}- g$  is the fluid velocity with respect
to $\Sigma_t$ ($-v$ is the celerity of $\Sigma_t$ with respect to
the fluid) and ${\mathbf{n}}^*$ is the transposed of
${\mathbf{n}}$.\\
To Eq. (\ref{GG})   of  motion, we have to add the equation of
balance of mass
\begin{equation}
{\partial \rho\over \partial t}\ +\ {\rm div} (\rho {\mathbf u}) =
0,  \label{K}
\end{equation}
Eq. (\ref{GH}) of conservation of the entropy and relation (\ref{J3}).\\
Due to the important fact that the first discontinuities of $\rho$
and $s$ are not existing in thermocapillary mixtures, weak
discontinuities of isentropic motions correspond to $\displaystyle
\rho,\ s,\ {\partial \rho /
\partial {\mathbf x}},\ {\partial s /\partial {\mathbf x}},$
continuous through the wave surfaces. As in \cite{hadamard}, we
denote by $[\,\,\, ]$ the jump of a tensorial quantity through a
surface of discontinuity $\Sigma_t$. The jump of a spatial
derivative of a continuous quantity  is colinear to the normal
vector to the wave surface; as in Hadamard's tensorial framework,
there exits two scalar fields $\lambda_1, \lambda_2$ on
$\Sigma_t$,
 such that
\begin{subequations}
\begin{eqnarray}
\bigg[\, {\partial \rho \over \partial {\mathbf z}}\,\bigg] = 0\
\Rightarrow\ \bigg[\, {\partial\over\partial {\mathbf z}}\,
({\partial \rho \over
\partial {\mathbf z}})^{^*}\,\bigg] = \lambda_1 {\mathbf N} {\mathbf N}^*\
\Rightarrow\ \bigg[\, {\partial\over\partial {\mathbf x}}\,
({\partial \rho \over
\partial {\mathbf x}})^{^*}\,\bigg] = \lambda_1  {\mathbf n} {\mathbf
n}^*,\ \ \ \
\label{M1}\\
\bigg[\, {\partial s \over \partial {\mathbf z}}\,\bigg] = 0\
\Rightarrow\ \bigg[\, {\partial\over\partial {\mathbf z}}\,
({\partial s \over
\partial {\mathbf z}})^{^*}\,\bigg] = \lambda_2 {\mathbf N} {\mathbf N}^*\
\Rightarrow\ \bigg[\, {\partial\over\partial {\mathbf x}}\,
({\partial s \over
\partial {\mathbf x}})^{^*}\,\bigg] = \lambda_2 {\mathbf n} {\mathbf n}^*, \ \ \ \
\label{M2}
\end{eqnarray}
\end{subequations}
and consequently
\begin{equation}
{\lambda_1 = [\, \Delta \rho\,]\ \ {\rm and} \ \ \lambda_2 = [\,
\Delta s\,] }. \label{M3}
\end{equation}
From $[{\mathbf V}] = 0$, we deduce $\displaystyle \Big[\,
{\partial {\mathbf V} \over
\partial {\mathbf z}}\,\Big] = {\mbox{{\boldmath
$\Xi$}}}\ {\mathbf N}^*$  with ${\mathbf N}^* = (-g,{\mathbf n}^*) $ and $ {\mbox{{\boldmath $\Xi$}}} =  \left(\begin{array}{c} 0 \\
{\mathbf H}
\end{array} \right)$ , where ${\mathbf H}$ is a 3-vector field on $\Sigma_t$.
Then,
$$
[\, {\mathbf \Gamma}] = \displaystyle \Big[\, {\partial {\mathbf
u} \over
\partial {\mathbf z}}\,\Big] {\mathbf V} =    {\mathbf H}\, {\mathbf N}^*\, {\mathbf V} \equiv v
\, {\mathbf H}.
$$
Equation of mass conservation (\ref{K}) is equivalent to
$$
{\partial \rho \over \partial {\mathbf z}}\, {\mathbf V} +\rho\,
{\rm Tr}\left({\partial {\mathbf V} \over
\partial {\mathbf z}}\right) = 0,
$$
where Tr denotes the trace operator. Then,
$\displaystyle\bigg[{\rm Tr}\left({\partial {\mathbf V} \over
\partial {\mathbf z}}\right)\bigg] = 0$ and  consequently,
\begin{equation}
{\mathbf n}^*\, {\mathbf H} = 0.  \label{N}
\end{equation}
Eq. (\ref{GH}) of conservation of entropy  implies
$$ {\partial \over \partial {\mathbf z}}({ds\over dt}) = 0
 \ \Leftrightarrow \ {\partial\over
\partial{\mathbf z}} \left({\partial{ s} \over
\partial {\mathbf z}}\, {\mathbf V}\right) = 0.
$$
Then,\ $\hskip 1.5 cm \displaystyle \Big [\, \left({\partial\over
\partial{\mathbf z}}  \Big({\partial{ s} \over
\partial {\mathbf z}}\Big)^*\,\right)^* {\mathbf V} +
 \left({\partial {\mathbf V} \over
\partial {\mathbf z}}\right)^* \left({\partial s\over
\partial {\mathbf z}}\right)^*\, \Big ] = 0.$\\
Due to the fact that $\displaystyle {\partial\over
\partial{\mathbf z}}  \Big({\partial{ s} \over
\partial {\mathbf z}}\Big)^*$ is a symmetric tensor,
$$
\lambda_2\, {\mathbf N} {\mathbf N}^*\, {\mathbf V} + {\mathbf
N}{\mbox{{\boldmath $\Xi$}}}^* \Big({\partial{ s} \over
\partial {\mathbf z}}\Big)^* = 0\ \  \Leftrightarrow\ \
{\mathbf N}\, \left (\lambda_2\, v + {\partial{ s} \over
\partial {\mathbf x}}\, {\mathbf H} \right) = 0
$$
or,
\begin{equation}
\lambda_2\, v + {\partial{ s} \over
\partial {\mathbf x}}\, {\mathbf H} = 0. \label{O}
\end{equation}
From Rankine-Hugoniot condition associated to Eq. (\ref{GG}), we
obtain the compatibility condition: $[-g \, \rho\, {\mathbf u}^* +
\rho\, {\mathbf n}^* {\mathbf u}{\mathbf u}^* - {\mathbf
n}^*\sigma] = 0,$ and the continuity of $ \rho, {\partial\rho/
\partial {\mathbf x}}, s, {\partial s/
\partial {\mathbf x}}$, yields $[\,{\rm div}\, {\mbox{{\boldmath
$\Phi$}}}\,] = 0 $ or $[\, C \Delta\rho + D\Delta s\,] = 0$ which
is equivalent to
\begin{equation}
 C\, \lambda_1 + D\,\lambda_2 = 0. \label{Q}
\end{equation}
Consequently, there exists a scalar field $\lambda_3$ on
$\Sigma_t$ such that
\begin{equation}
 \Big[\ {\partial\over\partial {\mathbf x}}\left( C \Delta\rho +
 D\Delta s\right)\,\Big] = \lambda_3\, {\mathbf n}^*.
  \label{R}
\end{equation}
Equation  of motion (\ref{G}) yields
$$
[\, {\mathbf \Gamma}\,] = [\,\theta\,]\, {\rm grad}\,s -
\lambda_3\, {\mathbf n},
$$
$$
 \ {\rm or}, \ \ \ \rho \,v \,  {\mathbf H} = - \left( D \lambda_1 +
E \lambda_2 \right)\, {\rm grad}\, s
  -\rho\, \lambda_3\, {\mathbf n}.
$$
By projection on the normal and tangent plane to $\Sigma_t$ and
taking relation (\ref{N}) into account, we obtain
\begin{equation}
  D\,({\mathbf n}^*\, {\rm grad} s)\,
\lambda_1 + E \,({\mathbf n}^*\, {\rm grad} s)\,
 \lambda_2   + \rho\,\lambda_3 = 0,,\label{R2}
\end{equation}
\begin{equation}
 \rho\, v\, {\mathbf H} =  - \left( D\,
 \lambda_1
+ E\,\lambda_2 \right)\, {\rm grad_{tg}} s ,  \label{S}
\end{equation}
where $ {\rm grad_{tg}} s$ denotes the tangential part of  $ {\rm
grad} s$  in $\Sigma_t$. Elimination of ${\mathbf H}$ in the
relation (\ref{S}) comes from relation (\ref{O}), and we get
\begin{equation}
D\, ({\rm grad_{tg}} s)^2\, \lambda_1 + \left(E\, ({\rm grad_{tg}}
s)^2 -\rho\, v^2\right)\, \lambda_2 =0.\label{R3}
\end{equation}
Consequently, we obtain the system (\ref{Q}), (\ref{R2}),
(\ref{R3}) of three homogeneous linear equations with respect to
the variables $\lambda_1, \lambda_2, \lambda_3$.
 The compatibility of the three equations
yields
\begin{equation}
 \rho\,  v^2  =  {(CE-D^2)\, ({\rm grad_{tg}} s)^2\over C}. \label{T}
\end{equation}
\subsection{Celerity of isentropic waves of acceleration}

\noindent The temperature in  liquid and vapor bulks is $T_0$.
Then relation (\ref{KK}) yields
\[
{\rm grad} \left(\,A (\rho-\rho_c)^2 + 2\, \rho\, s\, \right) =0\
\Leftrightarrow\ A (\rho-\rho_c)\,{\rm grad}\,\rho + {\rm
grad}\,\rho s = 0.
\]
The value $\rho=\rho_c$ of the matter density in the interface
corresponds to the maximum value of ${\rm grad}\,\rho$. The matter
density $\rho_c$ is characteristic of the interfacial matter. For
such a value, ${\rm grad}\,\rho s = 0$ and $s = (A^2/2B \rho_c)\,
(T_0-T_c)$.\\ Consequently, $\displaystyle
 {\rm grad}\,s = {A^2\over  2B \rho_c^2}\, (T_c-T_0)\,  {\rm
 grad}\,\rho,\  {\rm and}
$
$$
  v^2  =  {(CE-D^2)\, A^4\,
  ({\rm grad_{tg}} \rho)^2\, (T_c-T_0)^2\,\over {4\,C\,B^2\,
  \rho_c^5}}.
$$
 Due to relation (\ref{L}), we obtain when $\rho=\rho_c,\   $
 $
 \displaystyle C\, ({\rm grad_{tg}}\, \rho)^2 = {A^2\over 2\,B}\,
 (T_c-T_0)^2,
 $  and consequently,
\begin{equation}
  v^2  =  {(CE-D^2)\, A^6\,
   (T_c-T_0)^4\,\over {8\,C^2\,B^3\, \rho_c^5}}. \label{U}
\end{equation}
In the interfacial layer, ${\rm
 grad}\,\rho\,$ is normal to  iso-density surfaces.
 The isentropic waves of acceleration associated with a weak discontinuity
 shear the interfacial layer.
The wave celerity which is proportional to $(T_c-T_0)^2$, vanishes
at the critical point and can be calculated numerically by means
of a state equation.
\section{Results and discussion}
Near the critical point, for thermocapillary fluids, the variation
of matter density leads to the variations of entropy (see Eq.
(\ref{KK})). This extension of Cahn and Hilliard fluids seems at
first sight purely formal and at equilibrium yields  same results
as the classical van der Waals model does \cite{vdW}. In dynamics,
 motions in the normal direction to fluid
interfaces (as solitary waves) are not involved in an additive
dependance of the entropy gradient. In this paper, we see that the
dependance of entropy gradient is necessary for isentropic waves
of acceleration along the interfaces: the fact that the internal
energy depends not only on the gradient of matter density but also
on the gradient of entropy, yields a new kind of waves which does
not appear in the simpler models. It is easy to see they are
exceptional waves in the sense of Lax \cite{Ruggeri} and they
appear only in, at least, two-dimension spaces. Recent experiments
in space laboratories, for carbonic dioxide near its critical
point have showed the possibility of such waves \cite{Garrabos}
and should justify the well-founded of the thermocapillary fluid
model.


\begin{thebibliography}{99}

\bibitem{casal} {Casal P., Gouin H.},
Equations of motions of thermocapillary fluids, {Comptes Rendus
Acad. Sci. Paris} {306, II} {(1988)} {99-104}.

\bibitem{vdW} {van der Waals J.D.}, {Thermodynamique de la capillarit\'e
dans l'hypoth\`ese d'une variation continue de densit\'e,}
  {Archives N\'eerlandaises} {28}  {(1894-1895)} {121-209}.


\bibitem{cahn} {Cahn J.W.,
Hilliard  J.E.,} {Free energy of a non-uniform system III,}  {J.
Chem. Phys.} {31} {(1959)} {688-699}.

\bibitem{domb} {Domb C.,} {The critical point,}  {Taylor \& Francis,}
{London,} 1996.

\bibitem{slemrod} {Slemrod M.,} {Admissibility criteria for propagating
phase boundaries in a van der Waals fluid,} { Arch. Rat. Mech.
Anal.} {81} {(1983)} {301}.

\bibitem{trusk} {Truskinovsky L.,} Dynamics of non-equilibrium phase boundaries
 in a heat conducting non-linearly elastic medium {  P.M.M.} {51} {(1987)} {777-784}.

 \bibitem{rowlinson} {Rowlinson J.S., Widom B.,} {Molecular
theory of capillarity,}  {Clarendon Press, Oxford, 1984}.

\bibitem{gouin0} {Gouin H.,} {Thermodynamic form of the equation of motion for
perfect fluids of grade n,}  {Comptes Rendus Acad. Sci. Paris}
{305, II} {(1987)} {833-838}.

\bibitem{casal1} {Casal P., Gouin H.}, Non-isothermal liquid-vapour interfaces, { J. de
M\'ecanique Th\'eorique et Appliqu\'ee} {7} {(1988)} {689-718}.

\bibitem{gouin2} {Gouin H., Delhaye J.M.,} Material waves of a fluid
in the vicinity of the critical point  in {S. Morioka S.,
Wijngaarden L. (Eds.), Symposium on waves in liquid/gas and
liquid/vapor two-phase systems}, {Kluwer Publ., Netherlands,
1995}.


\bibitem{hadamard} {Hadamard J.,} {Le\c cons sur la propagation des ondes et
les \'equations de l'hydrodynamique,}  {Chelsea Pub., New York
1949}.


\bibitem{Ruggeri} {Boillat G.}, Non linear hyperbolic fields and waves {\it in} {Ruggeri T.  (Ed.)},
 {Recent mathematical methods
in nonlinear wave propagation}, Lecture Notes in Mathematics
 {1640},   Springer-verlag, Berlin, 1996.

\bibitem{Garrabos} {Garrabos Y.},
 {private communication}, 2003.

\end{thebibliography}
\end{document}